\documentclass[11pt]{article}
\usepackage{amsmath,amssymb,color,graphics,epsfig,cite}

\usepackage{amsfonts}

\newcommand{\hoch}[1]{$\, ^{#1}$}

\makeatletter
\@addtoreset{equation}{section}
\makeatother

\newcommand{\be}{\begin{equation}}
\newcommand{\ee}{\end{equation}}
\newcommand{\bea}{\setlength\arraycolsep{2pt} \begin{eqnarray}}
\newcommand{\eea}{\end{eqnarray}}
\newcommand{\nn}{\nonumber}

\def\ft#1#2{{\textstyle{\frac{\scriptstyle #1}{\scriptstyle #2} } }}
\def\fft#1#2{{\frac{#1}{#2}}}

\def\0{{\sst{(0)}}}
\def\1{{\sst{(1)}}}
\def\2{{\sst{(2)}}}
\def\3{{\sst{(3)}}}
\def\4{{\sst{(4)}}}
\def\5{{\sst{(5)}}}
\def\6{{\sst{(6)}}}
\def\7{{\sst{(7)}}}
\def\8{{\sst{(8)}}}
\def\sst#1{{\scriptscriptstyle #1}}

\def\del{{\partial}}

\def\cH{{{\cal H}}}
\def\im{{{\rm i\,}}}
\def\R{{\mathbb R}}

\begin{document}

\begin{flushright}
\hfill{MI-TH-1525}

\end{flushright}

\begin{center}
{\Large {\bf Generalised Smarr Formula and the Viscosity Bound for
Einstein-Maxwell-Dilaton Black Holes}}

\vspace{15pt}
{\bf Hai-Shan Liu\hoch{1,3}, H. L\"u\hoch{2} and C.N. Pope\hoch{3,4}}

\vspace{10pt}

\hoch{1} {\it Institute for Advanced Physics \& Mathematics,\\
Zhejiang University of Technology, Hangzhou 310023, China}

\vspace{10pt}

\hoch{2}{\it Center for Advanced Quantum Studies, Department of Physics, \\
Beijing Normal University, Beijing 100875, China}

\vspace{10pt}

\hoch{3} {\it George P. \& Cynthia Woods Mitchell  Institute
for Fundamental Physics and Astronomy,\\
Texas A\&M University, College Station, TX 77843, USA}

\vspace{10pt}

\hoch{4}{\it DAMTP, Centre for Mathematical Sciences,
 Cambridge University,\\  Wilberforce Road, Cambridge CB3 OWA, UK}

\vspace{20pt}

\underline{ABSTRACT}
\end{center}

   We study the shear viscosity to entropy ratio $\eta/S$ in the boundary
field theories dual to black hole backgrounds in theories of gravity coupled
to a scalar field, and generalisations including a Maxwell field and
non-minimal scalar couplings.  Motivated by the observation in simple
examples that the saturation of the $\eta/S\ge 1/(4\pi)$ bound is
correlated with the existence of a generalised Smarr relation for the
planar black-hole solutions, we investigate this in detail for the general
black-hole solutions in these theories, focusing especially on the
cases where the scalar field plays a non-trivial role and gives rise to
an additional parameter in the space of solutions. We find that a generalised
Smarr relation holds in all cases, and in fact it can be viewed as the
bulk gravity dual of the statement of the saturation of the viscosity to
entropy bound.  We obtain the generalised Smarr relation, whose existence
depends upon a scaling symmetry of the planar black-hole solutions,
 by two different but related methods, one based on integrating the
first law of thermodynamics, and the other based on the construction of
a conserved Noether charge.

\vfill {\footnotesize Emails: hsliu.zju@gmail.com \ \ \ mrhonglu@gmail.com\ \ \
pope@physics.tamu.edu}

\thispagestyle{empty}

\pagebreak

\tableofcontents
\addtocontents{toc}{\protect\setcounter{tocdepth}{2}}



\section{Introduction}

   The AdS/CFT correspondence \cite{adscft1,adscft2,adscft3} has provided many
remarkable insights into the connections between gravitational backgrounds in
string theory or more general settings, and strongly-coupled field theories
on the boundary of asymptotically anti-de Sitter spacetimes.  One of the
most striking results that has emerged is the intriguing universality of
the ratio $\eta/S=1/(4\pi)$
of the shear viscosity to entropy for wide classes of gauge theories
that are dual to gravitational backgrounds.  This led to the proposal
\cite{Policastro:2001yc,KSS,KSS0} of a universal bound $\eta/S\ge 1/(4\pi)$ for all materials. A number of papers have demonstrated the universality of this bound for a variety of supergravity and gravity theories \cite{Buchel:2003tz,Buchel:2004qq,Benincasa:2006fu,Landsteiner:2007bd}.
(See \cite{Cremonini:2011iq} for a review.)  It was shown in
\cite{Iqbal:2008by} that
the shear viscosity is determined by the effective coupling constant of the
transverse graviton on the horizon, by employing the membrane paradigm.
(This was confirmed by using the
Kubo formula in \cite{Cai:2008ph,Cai:2009zv}.)
In \cite{Brustein:2007jj}, it was shown that the black hole entropy is
determined by the effective Newtonian coupling
at the horizon, and that it is thus not surprising that the ratio of
the shear viscosity
to the entropy density is universal in the sense that the dependence
of the
quantity on the horizon is canceled.  Nevertheless, it naturally became
of interest
to seek counter-examples to the conjecture,
but within the framework of standard two-derivative field
theories, the bound seems to have been
remarkably robust. A violation was found, however, in
the result for  $\eta/S$ for the case of a bulk five-dimensional
theory of Einstein
gravity with a Gauss-Bonnet quadratic curvature correction \cite{Kats:2007mq,shenker}.
Non-universality also occurs in anisotropic configurations, where the
local rotational symmetry is broken
\cite{Natsuume:2010ky,Erdmenger:2010xm,Ovdat:2014ipa,Ge:2014aza}.  In this paper, we
shall focus only on isotropic configurations in two-derivative gravities.

  It is of interest to try to uncover some underlying understanding for
why the saturation of the $\eta/S\ge 1/(4\pi)$ is seemingly so widespread
in the classes of theories that have been studied.  A possible
line of thinking is suggested if we begin by looking
at the simple example of the calculation of $\eta$ for
the case of a planar Schwarzschild-AdS$_n$ black hole in pure Einstein
gravity with a cosmological constant, for which the metric is
\be
ds^2 = -\Big(g^2 r^2 -\fft{\mu}{r^{n-3}}\Big)\,dt^2 +
\Big(g^2 r^2 -\fft{\mu}{r^{n-3}}\Big)^{-1}\, dr^2 + r^2 \, dx^i dx^i\,.
\ee
One finds that
\be
 \eta= \fft{(n-1)\, M}{4\pi \,(n-2)\, T}\,,\label{eta0}
\ee
where $M= (n-2)\, \mu/(16\pi)$ is the mass per unit $(n-2)$-area and $T$
is the Hawking temperature.  From the scaling symmetry
\be
r = \lambda\, \hat r\,,\quad x^i = \lambda^{-1}\, \hat x^i\,,
\quad t =\lambda^{-1}\, \hat t\,,\quad
  \mu =\lambda^{n-1}\, \hat\mu
\ee
of the solution, and the scaling symmetries $T=\lambda\, \hat T$ and
$S=\lambda^{n-2}\, \hat S$ of the Hawking temperature and entropy,
 one has $M(\lambda^{n-2}\, \hat S)=\lambda^{n-1}\, \hat M(\hat S)$, and
hence acting with $\lambda\, \del/\del\lambda$
one can easily derive from the first law of thermodynamics $dM=TdS$ that
\be
M= \fft{n-2}{n-1}\, TS\label{gensmarr0}
\ee
for the planar Schwarzschild-AdS  solution.\footnote{This is a generalisation,
which works only for planar black holes,
of the well-known Smarr relation $M= [(n-2)/(n-3)]\, TS$ for
asymptotically-flat Schwarzschild black holes in $n$ dimensions.  (See
\cite{smarrpaper} for the original discussion of the Smarr formula in
four dimensions.) The
reason for the different coefficient of $TS$ is that a different scaling
symmetry arises in the asymptotically-flat case. We discuss the distinction
between the two types of Smarr relation in appendix A.}
Substituting (\ref{gensmarr0}) into (\ref{eta0}), we see that the result
$\eta/S=1/(4\pi)$ in this example can be attributed to the fact that
the generalised Smarr relation (\ref{gensmarr0}) holds for the
planar Schwarzschild black holes.

    In view of this observation in the simple example of the planar
Schwarzschild black holes, it is tempting to conjecture that the
universality of the viscosity to entropy ratio for the variety of
gravitational backgrounds that have been tested might be attributable to
the universal validity of the appropriate extension of the
generalised Smarr relation (\ref{gensmarr0}).

    It is not hard to check in some more complicated examples, such
as the case of charged planar black holes in Einstein-Maxwell theory
with a cosmological constant, that indeed the calculated $1/(4\pi)$ ratio
for the viscosity to entropy in this case is implied by the
generalised Smarr relation
\be
M= \fft{n-2}{n-1}\, (T S + \Phi\, Q)\,,
\ee
where $\Phi$ is the potential difference between
the horizon and infinity, and $Q$ is the conserved charge per unit area.

In this paper, our focus is on some rather more complicated examples
of theories admitting asymptotically-AdS black holes, in which a scalar
field with a scalar potential is present.  Our reason for considering such
cases is that one derivation of the generalised Smarr relation essentially
follows from thermodynamical considerations, and the thermodynamics of
black holes in these theories is quite subtle, and even somewhat
controversial.  We shall show that nevertheless, by following procedures
along the lines of those we described above in the simple example of
planar Schwarzschild-AdS black holes, we are able to derive generalised
Smarr relations that allow us to prove that the widely universal
$\eta/S= 1/(4\pi)$ result holds in these cases too.

  A rather remarkable aspect of the generalised Smarr relations we obtain is
the following.  The general planar black-hole solutions in the
Einstein-Scalar theories that we consider depend on two independent
parameters (mass, and what may loosely be called ``scalar charge''), but
these solutions seemingly cannot be constructed explicitly, on account
of the complexity of the equations.  One can construct the solutions
numerically, by setting initial data just outside the horizon and then
integrating the equations of motion out to infinity.  One can hence determine
the parameters of the asymptotic solution numerically in terms of the
parameters on the horizon, but this would appear on the face of it
to preclude the possibility
of obtaining an exact result for $\eta/S$, since this, as we have seen in
the simple example, is of the general form of $M$ (an asymptotic quantity)
divided by $TS$ (quantities defined on the horizon).  However, we find
that for the Einstein-Scalar black holes we can derive a generalised
Smarr formula that does precisely what is wanted, by providing an
exact formula expressing the mass in terms of the product $TS$.  Thus
although the expressions for the full set of asymptotic quantities in the
solutions can indeed only be found numerically in terms of the horizon
quantities, the precise one we need in order to calculate $\eta/S$ can be
calculated exactly.

  The bulk of this paper, therefore, is concerned with an exploration of the
generalised Smarr formula for planar black holes in certain theories
involving a scalar field in addition to gravity.  First, though, we begin
in section 2 with a derivation of the shear viscosity in the Einstein-Scalar
theories, by considering
transverse-traceless metric fluctuations around planar black-hole
backgrounds.
In section 3 we give a review of the
thermodynamics of black holes in Einstein-Scalar theories, and then
we use this in order to
derive the generalised Smarr relation that the planar black holes
satisfy.  In section 4 we combine the results of sections 2 and 3, to
show that we obtain the universal result $\eta/S= 1/(4\pi)$ in all these
examples, as a consequence of the universal validity of the generalised
Smarr relation.i\footnote{A rather different approach that relates the
$\eta/S= 1/(4\pi)$ result to universal features of black holes
was discussed in \cite{KSS0}, where it was shown to be related to the
low-energy graviton absoprtion  cross section.}
In section 5 we extend our results to planar black holes
in Einstein-Maxwell-Dilaton theories. This discussion includes the
added subtleties that arise in four dimensions, where non-trivial
dilatonic black holes carrying both electric and magnetic charge can arise.
In section 6 we extend the discussion further, by considering theories
where the scalar field couples non-minimally to gravity.  Section 7
contains a rather different derivation of the generalised Smarr relation
for the wide class of Einstein-Maxwell-Dilaton theories, where the scalar
couples minimally or non-minimally to gravity, based on the existence of
a conserved Noether charge in the planar black-hole solutions. We end with
conclusions in section 8.  In an appendix, we contrast the generalised
Smarr relation for planar asymptotically-AdS
black-hole solutions with the traditional Smarr relation for
asymptotically-flat spherically-symmetric black holes.

\section{Viscosity Bound in Einstein-Scalar Theories}

   In this section, we give a derivation of the $\eta/s$ ratio
for black holes in the theory of a scalar field minimally coupled
to gravity in a general dimension $n$, and with a general scalar potential
$V(\phi)$.  The theory is described by the $n$-dimensional Lagrangian
\be
e^{-1}\, {\cal L}_n = R - \ft12 (\del\phi)^2 - V(\phi)\,,\label{ESlag}
\ee
where $e=\sqrt{-\det g_{\mu\nu}}$.  We may take the $n$-dimensional
action to be
\be
S_n = \fft1{16\pi G}\, \int {\cal L}_n\, d^nx\,.\label{ndimaction}
\ee
The equations of motion are given by
\be
R_{\mu\nu}= \ft12\del_\mu\phi\, \del_\nu\phi + \fft{V}{n-2}\, g_{\mu\nu}\,,
\qquad \square\phi =\fft{\del V}{\del\phi}\,.\label{ESeom}
\ee

The black holes we shall consider, with flat spatial sections, take the
general form
\be
ds^2 = -h\, dt^2 + \fft{dr^2}{f} + r^2 \, dx^i dx^i\,,\label{metans}
\ee
where $h$ and $f$ are functions only of $r$.  Substituting into (\ref{ESeom})
gives the equations
\bea
\fft{V}{n-2} + \fft{f h'}{2rh} + \fft{f'}{2r} + \fft{(n-3)\, f}{r^2}
   &=& 0\,,\label{ESanseom1}\\
\fft{h''}{h} -\fft{{h'}^2}{2h^2} + \fft{h'\, f'}{2fh} -
       \fft{f'}{rf} + \fft{(n-3)\, h'}{rh} -\fft{2(n-3)}{r^2} &=&0\,,\label{ESanseom2}\\
\fft{(n-2)}{r}\, \Big( \fft{f'}{f}-\fft{h'}{h}\Big) &=& {\phi'}^2\label{ESanseom3}\,,\\
f\, \phi'' + \Big( \ft12 f' + \fft{f h'}{2h} + \fft{(n-2)\, f}{r}\Big)\,
   \phi' -\fft{\del V}{\del\phi} &=&0\,.\label{ESanseom4}
\eea

   We then consider a transverse-traceless metric perturbation in the
$(n-2)$-dimensional space of the spatial planar section, by making the
replacement
\be
dx^i dx^i \longrightarrow dx^i dx^i + 2\Psi\, dx^1 dx^2\,,\label{pert}
\ee
where for the present purposes it suffices to allow $\Psi$ to depend on
$r$ and $t$ only.  At the linearised level one finds, after making use
of the background equations (\ref{ESanseom1}-\ref{ESanseom4}), that $\Psi$ satisfies
\be
f\, \Psi'' +\Big[\fft{f h'}{2h} + \fft{(n-2)\, f}{r} +\ft12 f'\Big]\,
  \Psi' - \fft1{h}\, \ddot \Psi =0\,.\label{Psieqn}
\ee
For a perturbation of the form $\Psi(t,r) = e^{-\im\omega t}\, \psi(r)$,
we therefore have
\be
f\, \psi'' +\Big[\fft{f h'}{2h} + \fft{(n-2)\, f}{r} +\ft12 f'\Big]\,
  \psi' + \fft{\omega^2}{h}\, \psi =0\,.\label{perteq}
\ee

   If we consider a black hole solution of the equations (\ref{ESanseom1}-\ref{ESanseom4}),
with an horizon located at $r=r_0$, then near the horizon we shall have
the expansions
\be
h(r) = h_1\, [(r-r_0) + h_2\, (r-r_0)^2 +\cdots]\,,\qquad
f(r) = f_1\, (r-r_0) + f_2\, (r-r_0)^2 + \cdots\,.
\label{fhnear}
\ee
(We have written $h(r)$ with an overall scale $h_1$, which
is a ``trivial'' parameter, in the sense that it can be absorbed into a
rescaling of the time coordinate $t$.)
The equation (\ref{perteq}) therefore takes the form
\be
(r-r_0)^2\, f_1 h_1\, \psi'' + (r-r_0)\, f_1 h_1\, \psi'
   + \omega^2\, \psi \approx 0
\ee
near the horizon.  This can be solved exactly, leading to the
near-horizon ingoing solution
\be
\psi_{\rm in} \propto
\exp\Big[-\fft{\im\omega\, \log(r-r_0)}{\sqrt{f_1 h_1}}\Big]\,.
\label{psiin}
\ee
(The second, outgoing, solution is obtained by sending $\omega\longrightarrow
-\omega$ in (\ref{psiin}).)

   We assume that the black hole is asymptotic to AdS$_n$  with
$R_{\mu\nu}= -(n-1)\, g^2\, g_{\mu\nu}$ at infinity, where $g$ is the
inverse AdS$_n$ radius and so the scalar potential is such that
$V_\infty = V(\phi(\infty))= -(n-1)(n-2)\, g^2$.  Furthermore,
we shall focus on situations where the
metric functions $h$ and $f$ at large $r$ take the form
\be
h(r) = g^2 r^2 -\fft{\mu}{r^{n-3}} +\cdots\,,\qquad
f(r) = g^2 r^2 +\cdots\,.\label{fhlarge}
\ee
where the ellipses indicate terms with faster fall-offs than those preceding
them.  It is convenient then to rewrite the near-horizon
metric perturbation (\ref{psiin}) in the form
\be
\psi_{\rm in} = \exp\Big[-\fft{\im\omega}{\sqrt{f_1 h_1}}\,
                  \log\fft{h(r)}{g^2 r^2}\Big]\,.
\ee

   We may then seek the solution for the metric perturbation
away from the horizon, in the approximation where $\omega$ is small.
This suffices for the subsequent purpose of calculating the viscosity
in the boundary theory, since in the Kubo formula we need only
know $\psi$ up to linear order in $\omega$.  Making an ansatz of the
form
\be
\psi(r) =  \exp\Big[-\fft{\im\omega}{\sqrt{f_1 h_1}}\,
                  \log\fft{h(r)}{g^2 r^2}\Big]\, \Big(1-\im\omega U(r)\Big)\,,
\ee
and keeping terms only up to linear order in $\omega$, we find that
$U(r)$ satisfies the equation
\be
\fft{U''}{U'} = -\fft{n-2}{r} - \fft{(fh)'}{2fh}\,,
\ee
which can be solved to give
\be
U(r) = c_0 + c_1\, \int_{r_0}^r \fft{dr'}{{r'}^{n-2}\,
 \sqrt{f(r')\, h(r')}}\,.
\ee
In view of the near-horizon expansions (\ref{fhnear}), we see that
$U(r)$ would be logarithmically singular near $r=r_0$ unless we take
$c_1=0$.  Since we wish to normalise the metric perturbation so that
$\psi(r)\longrightarrow 1$ at $r=\infty$, we therefore conclude that
the required solution, valid to linear order in $\omega$, is simply
\be
\psi(r) = \exp\Big[-\fft{\im\omega}{\sqrt{f_1 h_1}}\,
                  \log\fft{h(r)}{g^2 r^2}\Big]\,.\label{psisol}
\ee
We see from this, and using (\ref{fhlarge}),
that at large $r$ we have the expansion
\be
\psi(r) = 1 + \fft{\im\omega\mu}{\sqrt{f_1 h_1}\, g^2\, r^{n-1}} +
 \cdots\,.
\ee
Bearing in mind that $f_1=f'(r_0)$ and $h_1=h'(r_0)$, and that
the Hawking temperature for black holes of the form
(\ref{metans}) is given by
\be
T= \fft{\sqrt{f'(r_0)\, h'(r_0)}}{4\pi}\,,
\ee
we see that the metric perturbation has the asymptotic form
\be
\psi(r) = 1+ \fft{\im\omega\mu}{4\pi g^2\, T}\, \fft1{r^{n-1}} + \cdots\,.
\label{psires}
\ee

   We can then calculate the viscosity using standard methods described
in the literature.  For our purposes, it is convenient to follow the
procedure given in \cite{sonsta,shenker}, making use of the Kubo formula.  The
first step involves calculating the terms in the action at
quadratic order in the metric perturbation $\Psi(t,r)$.  One should
include the Gibbons-Hawking term in the original action when doing this,
but the net effect of doing so is simply that the required quadratic action is
the one where second derivatives on $\Psi$ have been removed by
performing integrations by parts.  Thus we find the action at quadratic order
is given by
\be
S_n^{(2)} = \fft1{16\pi G}\int d^n x \Big[P_1 \, {\Psi'}^2 +
  P_2\, \Psi\, \Psi' + P_3\, \Psi^2 + P_4\, \dot\Psi^2\Big]\,,
\label{S2}
\ee
with
\bea
P_1&=&-\ft12 r^{n-2}\, \sqrt{fh}\,,\quad P_2=2r^{n-3}\, \sqrt{fh}\,,\nn\\
P_3&=& r^{n-4}\, \sqrt{fh}\, \Big(n-3+\fft{r (fh)'}{2fh}\Big)\,,\quad
P_4= \fft{r^{n-2}}{2\sqrt{fh}}\,.
\eea
The integrand in (\ref{S2}) can be written as
\be
\fft{d}{dr} \big(P_1\, \Psi\Psi' + \ft12 P_2 \Psi^2\big) +
   \fft{d}{dt}\, \big(P_4\, \Psi\dot\Psi\big) -
   \Psi\, \Big[ P_1\, \Psi'' + P_1'\, \Psi' + P_4\, \Psi''\Big]\,,
\ee
where we have used that $P_3- \ft12 P_2'=0$.  The last term, enclosed in
the square brackets, vanishes by virtue of the linearised equation
(\ref{Psieqn}) satisfied by $\Psi$, and so the integrand in (\ref{S2}) is
a total derivative.  The prescription described in \cite{sonsta,shenker}
requires knowing only the $P_1\, \Psi\Psi'$ term, for which we have
\be
\int {\cal L}_n^{(2)}\, dt d^{n-2}x =-
  \ft12\, r^{n-2}\, \sqrt{fh}\, \Psi\Psi'\Big|_{r=\infty} \,,\label{surfaction1}
\ee
Hence, from (\ref{psires}), we have
\be
\int {\cal L}_n^{(2)}\, dt d^{n-2}x =
\fft{\im\omega\mu(n-1)}{4\pi\, T}\,.
\ee
Using the prescription in \cite{sonsta,shenker}, we therefore find that
the viscosity is given by
\be
\eta =  \fft{(n-1)\, \mu}{64 \pi^2 T}\,. \label{etares}
\ee

It is worthwhile to emphasise at this point that the derivation we
have presented is valid for {\it any} planar black hole solution of
the equations of motion (\ref{ESeom}) that takes the form
(\ref{metans}).  That is to say, the result (\ref{etares}) was obtained without
needing to know the explicit form of the black hole solutions and
without needing to know the explicit form of the scalar potential
$V(\phi)$.  Thus we are able to calculate the viscosity for
the general two-parameter
black hole solutions to the
equations (\ref{ESanseom1}-\ref{ESanseom4}) explicitly, and for an arbitrary
choice of the scalar potential $V$, even though
the general two-parameter solutions can only be found numerically.

\section{Planar Black Hole Thermodynamics and Smarr Relation}

\subsection{Review of the thermodynamics of spherically-symmetric black holes}

   The thermodynamics of the general spherically-symmetric static
black hole solutions of the Einstein-Scalar theory described by
(\ref{ESlag}) has been discussed in \cite{Liu:2013gja,lupowe}.  These black
holes can be written in the form
\be
ds^2 = -h dt^2 + \fft{dr^2}{f} + r^2\, d\Omega_{n-2}^2\,,
\label{sphmet}
\ee
where $d\Omega_{n-2}^2$ is the metric on the unit $(n-2)$-sphere.  If we
assume the scalar potential $V(\phi)$ has a Taylor expansion around a
stationary point at $\phi=0$ of the form
\be
V(\phi) = -(n-1)(n-2) g^2 + \ft12 m^2\,\phi^2 + \gamma_3\, \phi^3 +
  \gamma_4\, \phi^4 + \cdots\,,\label{Vexp}
\ee
then $m$ is the mass of the scalar field $\phi$ in the asymptotically-AdS
background.  Defining
\be
\sigma=\sqrt{4\ell^2\, m^2 + (n-1)^2}\,,\label{sigmadef}
\ee
where $\ell=1/g$ is the AdS ``radius,'' one finds that the asymptotic
behaviour of the scalar field is of the form
\be
\phi(r) = \fft{\phi_1}{r^{(n-1-\sigma)/2}} + \fft{\phi_2}{r^{(n-1+\sigma)/2}}
   +\cdots\,.\label{phiinf}
\ee
The metric functions $h(r)$ and $f(r)$ have the asymptotic
forms\footnote{Depending on the value of the scalar mass parameter $m^2$,
there could be circumstances where there are terms in the large-$r$
expansion (\ref{phiinf}) have slower fall-offs
than those displayed explicitly.  In order to keep the discussion as
simple as possible, we shall postpone the discussion of such
cases until later.  The ``standard'' cases that we shall discuss first
occur when $0<\sigma<1$.}
\be
h(r)= g^2 r^2 + 1 -\fft{\mu}{r^{n-3}} + \cdots\,,\qquad
f(r) = g^2 r^2 + 1 + \cdots \,.\label{fhinf}
\ee
By substituting into the equations of motion, it is easy to see that the
large-$r$ expansion has three independent parameters, which we may take to
be $\mu$, $\phi_1$ and $\phi_2$.  All the remaining higher-order coefficients
are determined in terms of these.

   If we assume there is a black hole solution with an horizon at $r=r_0$,
the metric functions and scalar field will have near-horizon expansions
of the form
\bea
h(r)&=& h_1\,[(r-r_0) + h_2\, (r-r_0)^2 + h_3\, (r-r_0)^3 +\cdots]\,,\nn\\
f(r) &=& f_1\, (r-r_0) + f_2\, (r-r_0)^2 + f_3\, (r-r_0)^3 + \cdots\,,\nn\\
\phi(r) &=& \tilde\phi_0 + \tilde\phi_1\, (r-r_0) +
   \tilde\phi_2\, (r-r_0)^2 + \tilde\phi_3\, (r-r_0)^3 +\cdots\,.
\label{horexp}
\eea
(The tilded coefficients in the near-horizon
$\phi$ field expansion are not the same
as the expansion coefficients in the large-$r$ expansion (\ref{phiinf}).)
Note that $h_1$, which is common to all the terms in the
$h$ expansion, is a trivial parameter that is associated with the freedom
to rescale the $t$ coordinate.
By plugging the expansions into the equations of motion, one finds that
$h_1$ is, as expected, trivial and undetermined, and there are two
non-trivial parameters, which we may take to be $r_0$ and $\tilde\phi_0$.
All the other coefficients are then determined in terms of these.

  Although one cannot obtain the general black-hole solutions to the equations
of motion explicitly, it is easy to see by considering what happens if one
integrates out to infinity, starting from the family of near-horizon
solutions coming from (\ref{horexp}), which have two non-trivial parameters,
that all the members of the family will evolve in a non-singular fashion to
match on to members of the three-parameter family of asymptotic
solutions we discussed above.  Thus, we will find that the three
parameters in the asymptotic solutions are functions of the two non-trivial
parameters of the near-horizon solutions:
\be
\mu=\mu(r_0,\tilde\phi_0)\,,\qquad \phi_1=\phi_1(r_0,\tilde\phi_0)\,,
\qquad \phi_2=\phi_2(r_0,\tilde\phi_0)\,.
\ee
We can, if desired, view these parametric relations instead as saying
\be
\phi_2=\phi_2(\mu,\phi_1)\,,
\ee
so that we regard the asymptotic parameters $\mu$ and $\phi_1$ as the
independently-adjustable parameters of the general two-parameter
black-hole solutions, with $\phi_2$ determined as a function of $\mu$
and $\phi_1$.  Numerical calculations straightforwardly confirm the
existence of the two-parameter family of black-hole solutions.

  The fact that the full two-parameter family of black-hole solutions
cannot be constructed explicitly makes it a little difficult to give
a complete discussion of the black-hole dynamics or thermodynamics.
However, one can make some progress by applying Wald's analysis of
conserved charges associated with symmetries of the spacetime \cite{wald1,wald2}.
Applying
it to the timelike Killing vector $\del/\del t$, one derives the following
variations at infinity and on the horizon \cite{lupowe}:
\bea
\delta \cH_\infty &=& \fft{\omega_{n-2}}{16\pi}\,\Big[
   (n-2) \delta\mu +\delta K(\phi_1)  - \fft{\sigma\, g^2}{2(n-1)}\, [
(n-1-\sigma)\,\phi_1\, \delta\phi_2 -
              (n-1+\sigma)\, \phi_2\, \delta\phi_1]\Big] \,,\nn\\
\delta\cH_{r_0} &=& T \delta S\,,\label{deltaH}
\eea
where $\omega_{n-2}$ is the volume of the
unit $(n-2)$-sphere, $T=\kappa/(2\pi)$ is the Hawking temperature and
$S=\ft14 A$ is the Bekenstein-Hawking entropy.  The function $K(\phi_1)$ is
a polynomial in $\phi_1$, with coefficients
determined by the parameters in the Einstein-Scalar theory. For generic
values of the parameters $K(\phi_1)$ is zero, but for special values,
such as when $\sigma$ is an integer, $\log r$ terms generally occur
in the asymptotic expansions and their occurrence is associated with
$K(\phi_1)$ being non-zero \cite{lupowe}.
The Wald calculation
shows that $\delta\cH_\infty = \delta\cH_{r_0}$, and hence one has
\cite{lupowe}
\be
\fft{\omega_{n-2}}{16\pi}\,[(n-2) d\mu + dK]= T dS +
  \fft{\sigma\, g^2\, \omega_{n-2}}{32\pi (n-1)}\, [
(n-1-\sigma)\,\phi_1\, d\phi_2 -
              (n-1+\sigma)\, \phi_2\, d\phi_1] \,.\label{preFL1}
\ee

   Equation (\ref{preFL1}) provides a relation between the infinitesimal
variations of the parameters in the black-hole solutions. The left-hand side
is the infinitesimal variation of a quantity with the dimensions of energy,
or mass, and it is convenient to {\it define} that quantity to be the
{\it Thermodynamic Mass} of the black hole.  We shall write this as
$M_{\rm therm}$, defined by
\be
  M_{\rm therm}= \fft{\omega_{n-2}}{16\pi}\Big[(n-2)\, \mu + K\Big]
\,,\label{thermomass}
\ee
where $\mu$ is minus the coefficient of the $1/r^{n-3}$ term in the
large-$r$ expansion of $h(r)$.  Thus, in terms of the thermodynamic mass,
one has the first law
\be
dM_{\rm therm} = T dS +
  \fft{\sigma\, g^2\, \omega_{n-2}}{32\pi (n-1)}\, [
(n-1-\sigma)\,\phi_1\, d\phi_2 -
              (n-1+\sigma)\, \phi_2\, d\phi_1] \,.\label{FL1}
\ee

It should be emphasised that the thermodynamic mass $M_{\rm therm}$
is logically distinct
from the strict definition of the ``Hamiltonian Mass'' whose variation would
be given by the expression $\delta M_{\rm Ham}=\delta\cH_\infty$ if
$\delta\cH_\infty$ in (\ref{deltaH}) were integrable, which it is not for the
general two-parameter black-hole solutions.  The viewpoint
proposed in \cite{lupowe} is that rather than taking the non-integrability of
$\delta\cH_\infty$ to signal the end of any attempt to define a mass and
make use of the
relation (\ref{FL1}), it is more useful instead to interpret
(\ref{FL1}) as providing a {\it definition} of the ``thermodynamic
mass,'' which is then given by (\ref{thermomass}).  (See also \cite{Anabalon:2014fla,wen}.)
Thus the thermodynamic mass is an energy function whose variation is the
exact (and hence integrable) part of $\delta\cH_\infty$.  As we shall
see shortly, in the case of planar black holes (\ref{FL1}) may be employed
in order to derive a useful relation between the parameters of the
solutions.

   Large classes of explicit scalar hairy black-hole solutions
have been constructed in Einstein-Scalar theories in general dimensions
\cite{Anabalon:2012dw,Gonzalez:2013aca,Acena:2013jya,Feng:2013tza,Fan:2015tua,Faedo:2015jqa}.  However,
all these explicit solutions involve only one, rather than the
full complement of
two, parameters. It then follows, on dimensional grounds, that in these
solutions $\phi_1$ and $\phi_2$ are related by $\phi_2^{n-1-\sigma}=
c\, \phi_1^{n-1+\sigma}$, where $c$ is a dimensionless
constant that is independent of the parameter of the solutions.  As a
consequence, the differentials involving $d\phi_1$ and $d\phi_2$ in
(\ref{FL1}) cancel, and so the first law does not involve a contribution
from the scalar charge in these special solutions.
Numerical calculations confirm that two-parameter black-hole
solutions do exist \cite{Hertog:2004ns,lupowe}. For these solutions the
terms involving $\phi_1$ and $\phi_2$ do contribute non-trivially in the
first law (\ref{FL1}), and in fact their contribution in (\ref{FL1})
is essential in order for the right-hand side to be an exact form, and hence
integrable.  Numerical calculations in \cite{lupowe} confirmed that
(\ref{FL1}) is indeed obeyed for the general two-parameter solutions.

Finally we remark
that this technique of deriving the first law of thermodynamics have
been employed recently for AdS and Lifshitz black holes in a variety of
theories
involving Proca, Yang-Mills fields and higher-derivative curvature terms
\cite{Liu:2014tra,Liu:2014dva,Fan:2014ixa,Fan:2014ala}.

\subsection{The planar limit of the spherically-symmetric black holes}

   Having reviewed the essential points presented in \cite{lupowe}, we
now turn to the consideration of the thermodynamics of the
planar black holes we are studying in this paper.  The planar black holes
can be derived from those with spherically-symmetric spatial sections by
means of a limit procedure, in which we write the unit $S^{n-2}$
metric $d\Omega_{n-2}^2$  in (\ref{sphmet}) as
\be
d\Omega_{n-2}^2 = \fft{du^2}{1-u^2} + u^2 d\Omega_{n-3}^2\,,
\ee
define $u=k\, \bar u$, and then send $k$ to zero.  In the small-$k$
limit we have
\be
d\Omega_{n-2}^2 \longrightarrow k^2\, (d\bar u^2 + \bar u^2\,
d\Omega_{n-3}^2)\,,
\ee
which can be recognised as $k^2$ times the Euclidean metric in $(n-2)$
dimensions, written in hyperspherical polar coordinates.  One can then
make the standard transformation to Cartesian coordinates  $x^i$, so that we
have
\be
d\Omega_{n-2}^2 \longrightarrow k^2\, dx^i dx^i\,.
\ee
In order to keep the metric (\ref{sphmet}) non-singular in the limit when
$k$ goes to zero, we must define new barred radial and time coordinates:
\be
r= k^{-1}\, \bar r\,,\qquad t=k\, \bar t\,,\label{kscal0}
\ee
and make appropriate scalings of the various expansion coefficients in
the near-horizon and asymptotic forms for the metric and the scalar
field.  In particular, in the simplest situation of the ``standard''
cases that we are considering first, we shall have
\be
r_0=k^{-1}\, \bar r_0\,,\qquad \bar\mu = k^{1-n}\, \bar\mu\,,\qquad
\phi_1= k^{-(n-1-\sigma)/2}\, \bar\phi_1\,,\qquad
\phi_2= k^{-(n-1+\sigma)/2}\, \bar\phi_2\,.\label{kscal1}
\ee
After sending $k$ to zero, certain terms in the asymptotic
expansions of the metric functions and the scalar field scale away to
zero.  These terms include, but are not necessarily restricted to, the
``1'' terms in (\ref{fhinf}).  Dropping the bars after having taken $k$ to
zero, the metric functions now have the
expansions (\ref{fhlarge}), while the asymptotic expansion for the
scalar field will still take the same form as in (\ref{phiinf}). The
two-parameter family of black-hole solutions with spherical horizons
becomes a two-parameter family of planar black-hole solutions.  Later, we
shall discuss some more complicated situations where additional terms
disappear also when taking the planar limit.

\subsection{Thermodynamics and Smarr formula for planar black holes}

   An important new feature of the planar black holes is that they
have a scaling symmetry, absent in the spherical case, which means that
there exists a generalised Smarr formula relating the thermodynamic
mass $M$ to the other thermodynamic quantities.  In fact the scaling
symmetry is essentially a direct consequence of having derived the
planar black-hole solutions from the spherical ones by taking
the singular limit described in the previous subsection.
To study the generalised Smarr relation, we
first apply the planar limiting procedure discussed in the
previous subsection to
the first law (\ref{FL1}), in order to obtain the corresponding first
law for the planar black holes in the Einstein-Scalar theory. The
quantities $M$ and $S$ will now be viewed as mass and entropy densities,
by dividing out by the volume $\omega_{n-2}$ prior to taking the limit
when $k$ goes to zero.  It is then easy to see that the first law
(\ref{FL1}) becomes
\be
dM_{\rm therm} = T dS +
  \fft{\sigma\, g^2}{32\pi (n-1)}\, [
(n-1-\sigma)\,\phi_1\, d\phi_2 -
              (n-1+\sigma)\, \phi_2\, d\phi_1] \,,\label{FL0}
\ee
where the thermodynamic mass density is given by
\be
M_{\rm therm} = \fft{1}{16\pi}\, [(n-2)\,\mu + K(\phi_1)]\,.\label{Mdef}
\ee
We shall focus first on the generic cases where there are no $\log r$ terms
in the asymptotic expansions of the scalar and metric functions,
which means the function
$K(\phi_1)$ in the definition (\ref{Mdef}) of the thermodynamic
mass is absent.  However, it will be useful to define also what we shall
call the ``gravitational mass,'' which is simply given by
\be
M_{\rm grav} = \fft{(n-2)\, \mu}{16\pi}\,.\label{Mgravdef}
\ee
This is the ``naive'' mass that is simply associated with the coefficient
of the $1/r^{n-3}$ term in $g_{00}$, which is the leading-order term in the
asymptotic expansion of a massless spin-2 mode in the AdS background.
When there are no $\log r$ terms in the asymptotic expansions,
$M_{\rm therm}$ and $M_{\rm grav}$ are the same.  In cases where there are
$\log r$ terms, it will turn out that it is $M_{\rm grav}$ that appears
in the simplest form of the generalised Smarr relation.

   Proceeding for now with the generic discussion for the cases where there are
no $\log r$ terms,
it is easy to see that there exists a scaling symmetry under which
the coordinates transform as
\be
r=\lambda \hat r\,,\qquad x^i=\lambda^{-1}\, \hat x^i\,,\qquad
  t=\lambda^{-1}\, \hat t\,,\label{coordrescale}
\ee
with the parameters and thermodynamic quantities
correspondingly rescaling as
\bea
\mu &=& \lambda^{n-1}\, \hat\mu\,,\qquad \phi_1=\lambda^{(n-1-\sigma)/2}\,
\hat\phi_1\,,\qquad \phi_2=\lambda^{(n-1+\sigma)/2}\, \hat\phi_2\,,\nn\\
M_{\rm therm}  &=& \lambda^{n-1}\, \hat M_{\rm therm}\,,
\qquad T=\lambda\, \hat T\,,\qquad
  S=\lambda^{n-2}\, \hat S\,.\label{constrescale}
\eea
Whenever one has a scaling symmetry of this kind, it is always associated
with the existence of a generalised Smarr formula.  To derive the formula
in this example, it is useful as an intermediate step to define a
new energy function, which we shall call $E$, related to $M_{\rm therm}$
 by a Legendre
transformation such that on the right-hand side of the first law for $dE$
we have only the differentials $dS$ and $d\phi_1$.  Thus we define
\be
E = M_{\rm therm}
  -\fft{\sigma\, (n-1-\sigma)\, g^2}{32\pi (n-1)}\, \phi_1\, \phi_2\,,
\label{MErel}
\ee
in terms of which the first law (\ref{FL0}) becomes
\be
dE = TdS - \fft{\sigma\, g^2}{16\pi}\, \phi_2\, d\phi_1\,.
\label{EFL}
\ee
We may then view $E$ as a function only of $S$ and $\phi_1$, and
under the scaling symmetry we may deduce from $E=E(S,\phi_1)$ that
\be
E\big(\lambda^{n-2}\, \hat S,\lambda^{(n-1-\sigma)/2}\, \hat\phi_1\big)=
  \lambda^{n-1}\, \hat E(\hat S,\hat\phi_1)\,.\label{Escaling}
\ee
Acting with the Euler operator $\lambda \,\del/\del\lambda$ gives
\be
(n-2)\, \lambda^{n-2}\, \hat S\, \fft{\del E}{\del S} +
  \ft12(n-1-\sigma)\, \lambda^{(n-1-\sigma)/2}\, \hat\phi_1\,
   \fft{\del E}{\del\phi_1} = (n-1)\, \lambda^{n-1}\, \hat E\,.
\ee
Using (\ref{EFL}), we obtain the generalised Smarr relation\footnote{Note
that the scaling symmetry that is being used here is different from the
usual scaling, purely according to the ``engineering dimensions'' of
the parameters, that one uses when deriving the ``standard'' Smarr relation
for asymptotically-flat black holes.  Here, with $r\rightarrow \lambda\, r$,
we are scaling the mass parameter
$\mu$ so that the $g^2 r^2$ and the $-\mu/r^{n-3}$ terms in the metric
function $h$ both scale like $\lambda^2$, and hence $\mu\rightarrow
\lambda^{n-1}\, \mu$.  By contrast, in the usual Smarr relations for
asymptotically-flat black holes one scales $r\rightarrow \lambda\, r$ and
$\mu\rightarrow \lambda^{n-3}\, \mu$, so that the 1 and the $-\mu/r^{n-3}$
terms in the metric function $h$ are both scale-invariant.  This different
scaling accounts, in particular, for the different coefficient of the
$TS$ term in the generalised Smarr relation we have obtained here, in
comparison to the coefficient in the standard Smarr relation.}
\be
E= \fft{(n-2)}{(n-1)}\,  TS
 - \fft{\sigma\, (n-1-\sigma)\, g^2}{32\pi(n-1)}\, \phi_1\, \phi_2 \,.
\label{smarrE}
\ee
Written back in terms of the original energy function
$M_{\rm therm}$ using (\ref{MErel}),
we find the generalised Smarr relation
\be
M_{\rm therm}  = \fft{n-2}{n-1}\, TS\,.\label{smarrM0}
\ee
Since $M_{\rm therm}$ and $M_{\rm grav}$ are equal for the cases we have
discussed so far, where there are no $\log r$ terms in the asymptotic
expansions, we can also write the generalised Smarr relation as
\be
M_{\rm grav}  = \fft{n-2}{n-1}\, TS\,.\label{smarrM}
\ee
In fact, this way of writing the Smarr relation is preferable, since,
as we shall see in the next subsection, it is this relation, rather
than (\ref{smarrM0}), that holds in the cases where there are $\log r$
terms in the asymptotic expansions and hence when $M_{\rm therm}$ and
$M_{\rm grav}$ are unequal.

   It should be emphasised that even though the first law (\ref{FL0})
for the general two-parameter planar black hole
solutions involves the variations of the scalar parameters $\phi_1$ and
$\phi_2$, it has turned out that the generalised Smarr relation (\ref{smarrM})
involves only the product $TS$, with a zero coefficient
for the term involving the
product $\phi_1\, \phi_2$ that one might \`a priori have expected.  Of
course, the Smarr relation for a Legendre-transformed energy, such as the
quantity $E$ we defined in (\ref{MErel}), {\it does} then have a
$\phi_1\, \phi_2$ term, as seen in eqn (\ref{smarrE}).

\subsection{$\log r$ terms and anomalous scaling}

   In the previous section, we gave a rather general analysis of the
derivation of the generalised Smarr relation (\ref{smarrM})
obeyed by the gravitational
mass $M_{\rm grav}=(n-2)\mu/(16\pi)$ in the case of planar black holes.
In certain
cases, depending upon the value of the parameter sigma defined in
(\ref{sigmadef}), and upon specific features in the
scalar potential, $\log r$ terms can arise in the large-$r$
asymptotic expansions
of the metric functions and the scalar field.  This leads to modifications
in the scaling argument that we used previously in deriving the generalised
Smarr relation.

   A nice illustrative
example is provided by the case of $\sigma=1$ in $n=4$ dimensions.  From
(\ref{sigmadef}), it can be seen that this corresponds to $m^2=-2g^2$,
which is precisely the value of $m^2$ that arises for the scalar fields
in gauged four-dimensional supergravity in the maximally symmetric
${\cal N}=8$ AdS$_4$ background.  The asymptotic expansions of the
scalar and metric functions were presented for this example for spherical
black holes in \cite{lupowe}:
\bea
\phi &=& \fft{\phi_1}{r} + \fft{\phi_2}{r^2} -
   \fft{3\gamma_3\, \phi_1^2 \log r}{g^2 r^2} +\cdots\,,\nn\\
h &=& g^2 r^2 + 1 -\fft{\mu}{r} +\cdots\,,\nn\\
f &=& g^2 r^2 +1  +\ft14 g^2 \phi_1^2 +\fft{f_1}{r} -
    \fft{2\gamma_3\, \phi_1^3\, \log r}{r} +\cdots\,,\label{sphcase}
\eea
where $f_1= -\mu +\ft13\gamma_3\, \phi_1^3 +\ft23 g^2\, \phi_1\, \phi_2$.
Note that here $\gamma_3$ is the coefficient of the cubic term in
the Taylor expansion (\ref{Vexp}) of the scalar potential.  In ordinary
${\cal N}=8$ gauged supergravity $\gamma_3$ vanishes, and in that case
no $\log r$ terms are present in the asymptotic expansions.  But more
generally, we may consider theories where $\gamma_3\ne 0$.  In fact this
situation can arise in the recently-discovered $\omega$-deformed
${\cal N}=8$ gauged supergravities \cite{DalLAgata:2012bb,dewitnicnew}.

  Because of the presence of the $\log r$ terms, we must modify the
scaling transformations given in (\ref{kscal0}) and (\ref{kscal1}) before
taking the $k\rightarrow 0$ limit to get the planar black hole
solution.  Specifically, the coordinate scalings in (\ref{kscal0}) are
unchanged, but the transformation for $\phi_2$ is modified, so that
(\ref{kscal1}) becomes, for this $n=4$, $\sigma=1$ case,
\be
r_0=k^{-1}\, \bar r_0\,,\qquad \bar\mu = k^{-3}\, \bar\mu\,,\qquad
\phi_1= k^{-1}\, \bar\phi_1\,,\qquad
\phi_2= k^{-2}\, \bar\phi_2 - 3\gamma_3\,\bar\phi_1^2\, k^{-2} \,\log k
\,.\label{kscal1mod}
\ee
The net effect, after sending $k$ to zero, and
up to the orders displayed in (\ref{sphcase}), is
that the ``1'' terms in the metric functions disappear, and so for the
$n=4$, $\sigma=1$ planar black holes we have
\bea
\phi &=& \fft{\phi_1}{r} + \fft{\phi_2}{r^2} -
   \fft{3\gamma_3\, \phi_1^2 \log r}{g^2 r^2} +\cdots\,,\nn\\
h &=& g^2 r^2  -\fft{\mu}{r} +\cdots\,,\nn\\
f &=& g^2 r^2  +\ft14 g^2 \phi_1^2 +\fft{f_1}{r} -
    \fft{2\gamma_3\, \phi_1^3\, \log r}{r} +\cdots\,,\label{planarcase}
\eea
After taking the $k\rightarrow 0$ limit the first law, calculated in
\cite{lupowe}, takes the form
\be
dM_{\rm therm} = TdS +
  \fft{g^2}{48\pi}\, (\phi_1\, d\phi_2 - 2\phi_2\, d\phi_1) \,,
\ee
with
\be
M_{\rm therm} = \fft{1}{16\pi}\, (2\mu + K(\phi_1))\,,\qquad
K(\phi_1) = \ft13 \gamma_3\, \phi_1^3\,.\label{Mtherm41}
\ee
The presence of the $K(\phi_1)$ term here is associated with the occurrence
of the $\log r$ terms in the asymptotic expansions of the scalar and metric
functions.

  Turning now to the derivation of the generalised Smarr relation for this
example, the previous expressions for the
rescalings given in (\ref{coordrescale} and (\ref{constrescale})
also require modification because of $\log r$ terms.
The coordinate rescalings themselves are unchanged,
but now we find that in order for $\phi(r)$ to be invariant, and for
$h(r)$ and $f(r)$ to scale with overall $\lambda^2$ factors as they did
before, we must now have, in this $n=4$ and $\sigma=1$ case, that
\be
\mu =\lambda^3\,  \hat\mu\,,\qquad \phi_1=\lambda\, \hat\phi_1\,,\qquad
\phi_2 = \lambda^2\, \hat\phi_2 +
            3\gamma_3\, g^{-2}\, \hat\phi_1^2\, \lambda^2\, \log\lambda\,.
\label{anomscal}
\ee
Note that we still have $M_{\rm therm} =\lambda^3\,
\hat M_{\rm therm}$.  Defining the Legendre-transformed energy
function $E$ as before using (\ref{MErel}), we have
\be
E= M -\fft{g^2}{48\pi}\, \phi_1\, \phi_2\,,\label{Edef1}
\ee
which therefore obeys the first law
\be
dE = T dS -\fft{g^2}{16\pi}\, \phi_2\, d\phi_1 \,,\label{EFLnew}
\ee
and so $E$ can again be treated as a function of $S$ and $\phi_1$.
It is important to note that although $M_{\rm therm}$ scales in
the standard way, $M_{\rm therm}=\lambda^3\, \hat M_{\rm therm}$,
it follows from (\ref{anomscal}) and (\ref{Edef1}) that $E$ obeys the
anomalous scaling transformation
\be
E = \lambda^3\, \hat E -
 \fft{\gamma_3}{16\pi}\, \lambda^3\, \hat\phi_1^3\, \lambda^3\, \log\lambda
\,.
\ee
This leads to a modification in the standard scaling relation (\ref{Escaling}),
leading, in this $n=4$, $\sigma=1$ case, to
\be
E(\lambda^2\, \hat S,\lambda\, \hat \phi_1) =
        \lambda^3\, \hat E -
  \fft{\gamma_3}{16\pi}\, \lambda^3\, \hat\phi_1^3\, \lambda^3\, \log\lambda
\,.
\ee
Acting with the Euler operator $\lambda\, \del/\del\lambda$ and using
the first law (\ref{EFLnew}), we conclude that $E$ obeys
\be
E = \ft23 TS - \fft{g^2}{48\pi}\, \phi_1\, \phi_2+\fft{\gamma_3}{48\pi}\,
\phi_1^3\,.
\ee
Returning now to the thermodynamic mass $M_{\rm therm}$,
related to $E$ by (\ref{Edef1}),
we therefore obtain the generalised Smarr relation
\be
M_{\rm therm} = \ft23 TS +\fft{\gamma_3}{48\pi}\, \phi_1^3\,.
\ee
Finally, we see from definitions (\ref{Mgravdef}) and (\ref{Mtherm41})
that $M_{\rm grav}$ obeys the very simple generalised Smarr relation
\be
M_{\rm grav} = \ft23 TS\,.
\ee
Thus, despite the occurrence of the unusual new feature of $\log r$ terms in
the asymptotic expansions, and the associated anomalous scaling law for
$\phi_2$ as seen in (\ref{anomscal}), we find that in the end the
``gravitational  mass'' $M_{\rm grav}=\mu/(8\pi) $
continues to satisfy the same generalised
Smarr relation (\ref{smarrM}) (specialised to $n=4$) as in the
standard cases with no anomalous scaling.

   The phenomenon we have illustrated above for the special case of
$n=4$ and $\sigma=1$ is representative of what happens in all cases
where planar black holes solutions in the Einstein-Scalar theory
have $\log r$ terms in the asymptotic expansions for the scalar and metric
functions.  The conclusion in all cases is that the generalised Smarr
relation (\ref{smarrM}) holds, with the gravitational mass defined in
(\ref{Mgravdef}) appearing on the left-hand side.  A few further examples,
for black holes with spherical spatial sections, can be found in
\cite{lupowe}.  By applying the appropriate scalings, analogous to
(\ref{kscal0}) and (\ref{kscal1mod}), one can obtain the corresponding
planar black holes solutions in the limit when $k$ goes to zero.  The
derivation of the generalised Smarr relation then goes in a manner that
is very analogous to our derivation above, leading in general to the
conclusion that $M_{\rm grav}$ is related to $T$ and $S$ by
(\ref{smarrM}).  A further point worth noting is that in some
cases, the effect of taking the $k\rightarrow 0$ planar limit can be
to remove more than just the ``1'' terms in the asymptotic expansions of
the metric functions $h$ and $f$.  In particular, one finds that in the
metric function $h$, whenever the spherical black hole solutions have
terms with $r$ dependence lying between the leading-order $g^2 r^2$ term
and the mass term $-\mu/r^{n-3}$, then these terms scale away to zero when
$k$ goes to zero.  Thus, for the planar black holes one always finds that
the asymptotic expansions of the functions $h$ and $f$ are of the form
given in (\ref{fhlarge}), with no intermediate powers between $g^2 r^2$
and $-\mu/r^{n-3}$ in $h$, even if there are such intermediate powers in the
corresponding spherical black-hole solutions.  We give a
proof of this statement in section 7.

\subsection{Imaginary $\sigma=\im\, \tilde\sigma$}

   We have established at this point that for all real values of the
constant $\sigma$, defined in terms of the mass $m$ of the scalar field
by (\ref{sigmadef}), there is a generalised Smarr formula given by
(\ref{smarrM}), relating the ``gravitational mass'' $M_{\rm grav}$ defined
in (\ref{Mgravdef}) to the product of the entropy and temperature of the
black hole.  By a further extension of these arguments, we can include also
the case where $m^2<-\ft14 g^2\, (n-1)^2$, corresponding to a scalar field
whose mass-squared is more negative than the Breitenl\"ohner-Freedman
bound.  As formal solutions of the equations, Einstein-Scalar black holes
exist also in this regime where $\sigma$ is imaginary.  They are discussed
in \cite{lupowe} for the case of spherically-symmetric black holes.

   Before taking the scaling limit to obtain the planar black holes,
it is helpful to rewrite the spherically-symmetric
black holes with $\sigma=\im\tilde\sigma$ that are discussed in \cite{lupowe}
in terms of a reparameterisation of the two scalar field coefficients
$\phi_1$ and $\phi_2$, by writing
\be
\phi_1= \Phi\, \sin\chi\,,\qquad \phi_2= \Phi\, \cos\chi\,.
\ee
The asymptotic expansions presented in \cite{lupowe} then become
\bea
\phi &=& \fft{\Phi\, \sin(\chi + \ft12\tilde\sigma\, \log r)}{r^{(n-1)/2}}
   +\cdots\,,\qquad
h = g^2 r^2 + 1 - \fft{\mu}{r^{n-3}} +\cdots\,,\nn\\
f &=& g^2 r^2 + 1 + \fft{1}{r^{n-3}} \Big[-\mu \label{imagexp}\\
&&+\fft{g^2\, \Phi^2\,
\big( 2(n-1)^2 \,\sin^2(\chi + \ft12\tilde\sigma\, \log r) +
 \tilde\sigma^2 - (n-1)\,\tilde\sigma\, \sin(2\chi+\tilde\sigma\, \log r)
\big)}{8(n-1)(n-2)} \Big] +\cdots\,.\nn
\eea
The first law obtained in \cite{lupowe} becomes
\be
dM_{\rm therm} = T dS +\fft{g^2\, \tilde\sigma\, \omega_{n-2}}{16\pi}\,
                \Phi^2\, d\chi\,,
\ee
with
\be
M_{\rm therm}= \fft{\omega_{n-2}}{16\pi}\, \Big[ (n-2)\,\mu + K\Big]\,,
\qquad K= -\fft{g^2\, \tilde\sigma^2}{8(n-1)}\, \Phi^2\,.\label{imagM}
\ee

    Applying the scalings (\ref{kscal0}), together with
\be
\mu=k^{1-n}\, \bar\mu\,,\qquad \Phi=k^{-(n-1)/2}\, \bar\Phi\,,\qquad
\chi = \bar \chi +\ft12 \tilde\sigma\, \log k\,,
\ee
we obtain the planar limit of these black holes.  The effect in
the expansions displayed in (\ref{imagexp}) is to remove the ``1'' terms
in the expressions for $h$ and $f$.  The first law and the
expression for $M_{\rm therm}$ are unchanged, except that we now, as usual,
omit the factors $\omega_{n-2}$.  The scaling symmetry of the planar
black holes is given by
\be
r=\lambda\, \hat r\,,\qquad \mu=\lambda^{n-1}\, \hat \mu\,,\qquad
\Phi=\lambda^{(n-1)/2}\, \hat\Phi\,,\qquad
\chi = \hat\chi -\ft12\tilde\sigma\, \log\lambda\,,
\ee
with $S=\lambda^{n-2}\, \hat S$ and $T=\lambda \, T$ as usual, and so
applying the Euler operator $\lambda\, \del/\del\lambda$ to
\be
M_{\rm therm}(\lambda^{n-2}\, \hat S,
\hat \chi -\ft12\tilde\sigma\,\log\lambda) =\lambda^{n-1}\,
  \hat M_{\rm therm}(\hat S,\hat\chi)
\ee
and
using the first law, we obtain
\be
M_{\rm therm} = \fft{(n-2)}{(n-1)}\, TS -
   \fft{g^2\, \tilde\sigma^2}{126(n-1)\, \pi}\, \Phi^2\,.
\ee
It then follows from (\ref{Mgravdef}) and (\ref{imagM}) that the
gravitational mass $M_{\rm grav}$ again satisfies the simple
generalised Smarr relation (\ref{smarrM}).

   This completes our derivation of the generalised Smarr relation
(\ref{smarrM}) for the general two-parameter planar black hole
solutions in the Einstein-Scalar theories described by the Lagrangian
(\ref{ESlag}).  It was not necessary, for the purpose of this derivation,
to know the
explicit form of these solutions, nor even to know the detailed form of the
scalar potential.  It is rather remarkable that nonetheless, we are able
to obtain an exact expression for the coefficient $\mu$ of the ``mass term''
$-\mu/r^{n-3}$ in the large-$r$ expansion of the metric function $h(r)$,
purely in terms of $T$ and $S$, which are quantities defined on the
horizon of the black hole.  As we shall discuss in the next section, this
allows us to calculate the exact viscosity to entropy ratio for all
the Einstein-Scalar black holes, even though it is only possible to
construct the actual black-hole solutions numerically.

\section{Viscosity Ratio for the Einstein-Scalar Planar Black Holes}

   We are now in a position to give an explicit evaluation of the
ratio $\eta/S$ for the general two-parameter family of Einstein-Scalar black
holes.  From the expression (\ref{etares}) for the viscosity, we see that
\be
\fft{\eta}{S} = \fft{(n-1)\, \mu}{64\pi^2\, T S}\,.
\ee
 From the definition (\ref{Mdef}) of the gravitational mass, we then find
\be
\fft{\eta}{S} = \fft{(n-1)}{(n-2)} \fft{M_{\rm grav}}{TS}\, \fft1{4\pi}\,.
\ee
Finally, it follows from the generalised Smarr relation (\ref{smarrM}) that
\be
  \fft{\eta}{S} = \fft1{4\pi}\label{ratio}
\ee
for the entire two-parameter family of planar black holes in the
Einstein-Scalar theory described by (\ref{ESlag}).

   The intriguing conclusion from this discussion is that for the
entire two-parameter family of planar black hole solutions in Einstein-Scalar
gravity, the universality of the $1/(4\pi)$ viscosity to entropy ratio
can be seen to be due to the universal validity of the generalised
Smarr relation (\ref{smarrM}).  Furthermore, it was not necessary to be
able to construct the solutions explicitly (and indeed, the general
solution can {\it only} be obtained numerically), and the universal result
holds regardless of the detailed form of the scalar potential.

Large classes of explicit scalar hairy planar black-hole solutions
have been constructed in Einstein-Scalar theories in general dimensions
\cite{Acena:2013jya,Feng:2013tza,Fan:2015tua}, and it can easily be verified
that they indeed all satisfy the generalised Smarr relation.  However,
since all these explicit solutions involve only one, rather than the
general two, parameters, it follows from the earlier discussion in the end of section 3.1 that
the first law does not involve a contribution from the scalar charges.
The fact that the generalised Smarr relation holds, and the
consequent saturation of the viscosity bound, is then rather
straightforward and not especially remarkable in the case of these
special solutions. Indeed an explicit demonstration of $\eta/S=1/(4\pi)$
for such a one-parameter family of scalar black holes was
performed in \cite{Roychowdhury:2015cta}. Numerical black-hole
solutions with the complete complement of two independent parameters
have been constructed in \cite{Hertog:2004ns,lupowe}.  We have confirmed for
these solutions, and demonstrated numerically, that the generalised
Smarr formula is indeed obeyed for these general solutions.
Since we shall in any case present an alternative proof of the
generalised Smarr relation, in section 7, we shall not present the numerical
calculations here.

\section{Einstein-Maxwell-Dilaton Theory}

\subsection{General result}

We now consider the Einstein-Scalar theory coupled to a Maxwell field
described by the 1-form potential $A$.  A general class of Lagrangians is
given by
\be
e^{-1}{\cal L} = R- \ft12 (\partial\phi)^2 - \ft14 Z(\phi) F^2 - V(\phi)\,,
\label{emdlag}
\ee
where $F=dA$.  In supergravities, $Z$ is typically an exponential
function of $\phi$, but here we shall allow $Z$ to be an arbitrary
function of $\phi$.  Exact solutions of charged black holes with certain more general $Z$ are constructed in \cite{fanlu}.
 We shall assume for convenience that, as in the Einstein-Scalar case,
the relevant stationary point of the scalar potential is at $\phi=0$, so
that $\phi$ will asymptotically approach zero at $r=\infty$.
As in the pure Einstein-Scalar case we discussed
previously, we shall consider  planar black hole solutions, taking the
form
\be
ds_5= - h(r) dt^2 + \fft{dr^2}{f(r)} + r^2\,dx^i dx^i\,,\qquad
A = a(r)\,  dt\,,\qquad \phi=\phi(r)\,.\label{chargedbb}
\ee
The Maxwell equation implies that
\be
a'=\fft{q}{Z\,r^{n-1}} \sqrt{\fft{h}f}\,,\label{aprime}
\ee
where the $q$ is the electric charge (density) parameter.
The conserved electric
charge density is given by
\be
Q_e = \fft{1}{16\pi \omega_{n-2}} \int_{r\rightarrow \infty} Z {*F}
= \fft{q}{16\pi}\,.
\ee

   As in the Einstein-Scalar case, we consider a transverse-traceless metric
perturbation in the $(n-2)$-dimensional space of the spatial planar
section, by making the replacement (\ref{pert}).  At linearised order,
the perturbation can again be solved straightforwardly.  Taking $\Psi(t,r)=
e^{-\im\omega t}\, \psi(r)$, we find
\be
\psi(r)=
\exp\Big[-\fft{\im \omega}{\sqrt{h_1 f_1}}
\log \fft{h}{g^2 r^2}\Big]
(1 - \im \omega U + {\cal O}(\omega^2))\,,\label{chargedpsi}
\ee
where
\be
U=c_0 - \fft{q}{\sqrt{h_1f_1}}\int\fft{a-c_1}{r^{n-2}\sqrt{h f}}\,,
\ee
and $f_1$ and $h_1$ are the coefficients of the leading-order terms in the
near-horizon expansions (\ref{fhnear}).
Making a gauge choice so that $a(r)$ vanishes on the horizon, we must
then require
that the two integration constants $c_1$ and $c_0$ both vanish, so that $U$
is non-singular on the horizon and that $\Psi$ equals 1 at $r=\infty$.

We shall consider black holes that are asymptotic to AdS$_n$, with
\be
h\sim g^2r^2 - \fft{\mu}{r^{n-3}} + \cdots\,,
\ee
It follows that
\bea
\psi &=& 1 + \fft{\im \omega}{g^2 r^{n-1}}\,
\fft{\mu - \fft{1}{n-1}\Phi_e q}{4\pi T} + {\cal O}(\omega^2)\cr
&=& 1 + \fft{\im \omega}{g^2 r^{n-1}}
\fft{4}{T} \Big(\fft{M_{\rm grav}}{n-2}
-\fft{\Phi_e Q_e}{n-1}\Big) + {\cal O}(\omega^2)\,,
\eea
where $M_{\rm grav}$ is defined in (\ref{Mgravdef}) and
$\Phi_e=-a(\infty)$ is the electric potential at infinity.
The relevant part of the surface term of the action for the
linear mode $\Psi$ is given by (\ref{surfaction1}) and hence the
viscosity/entropy ratio is given by
\be
\fft{\eta}{S} = \fft{1}{4\pi\, T S} \Big(\fft{n-1}{n-2} M_{\rm grav}
         - \Phi_e Q_e\Big)\,.\label{chargedetas}
\ee

    The general planar black-hole solution in $n\ge 5$ dimensions
involves three independent parameters, namely the mass,
the scalar charge and the electric charge.  The first law of thermodynamics
is given by
\be
d M_{\rm therm} = T dS + \Phi_e dQ_e +
  \fft{\sigma\, g^2}{32\pi (n-1)}\, [
(n-1-\sigma)\,\phi_1\, d\phi_2 -
              (n-1+\sigma)\, \phi_2\, d\phi_1] \,,
\ee
where the scalar contribution and the relation between $M_{\rm therm}$
and $M_{\rm grav}$ was extensively discussed  in the previous
sections. Since the scaling behavior of $(\Phi_e,Q_e)$ is the same
as that of $(T,S)$, it follows from a straightforward extension of
the earlier discussion
that the generalised Smarr formula will
given by
\be
M_{\rm grav}=\fft{n-2}{n-1} \, (T S + \Phi_e Q_e)\,,\label{smarrPhi}
\ee
Thus the viscosity/entropy ratio is again given by (\ref{ratio}).

    In four dimensions, the Maxwell field $A$ can carry both electric
and magnetic charges, with the gauge potential given by
\be
A=\fft{q}{Z\,r} \sqrt{\fft{h}{f}}\, dt + p x_1 dx_2\,.
\ee
Note that there is no continuous electric/magnetic duality symmetry that
would allow one to rotate the system into a purely electric or purely magnetic
complexion, except in the special case when $Z(\phi)$ is just a constant.  Thus
the electric and magnetic charge parameters are genuinely independent
parameters in the solutions, except in the $Z=\hbox{constant}$ special case.
The electric and magnetic charge densities are now
\be
Q_e = \fft{q}{16\pi}\,,\qquad Q_m = \fft{p}{16\pi}\,.
\ee

   The derivations of the first law and the generalised Smarr relation proceed
in close analogy to the previous case we discussed, and the upshot is that
the four-parameter planar black holes obey the relation
\be
M_{\rm grav}=\fft{n-2}{n-1}\,
(T S + \Phi_e Q_e + \Phi_m\, Q_m)\,,\label{n4smarr}
\ee

For the linearised transverse and traceless mode
$\Psi(r,t)=e^{-{\rm i}\omega t} \, \psi(r)$, we find that up to linear order
in $\omega$,
\be
\psi= \exp\Big[-\fft{{\rm i}\omega}{\sqrt{h_1 f_1}}
\log \fft{h}{g^2 r^2}\Big]\,  (1 - {\rm i} \omega U)\,,
\ee
where
\be
U=c_0 - \fft{1}{\sqrt{h_1 f_1}}\int dr \fft{W}{r^2\sqrt{hf}}\,,\qquad
W'=\fft{1}{r^2} \Big(\fft{q^2}{Z} + p^2 Z\Big) \sqrt{\fft{h}{f}}\,.
\ee
Note that the $q^2$ term above is given by $q a'$, and the
corresponding term becomes $\Phi_e q$.  Owing to the electric and
magnetic duality, the remaining term in $W$ gives $\Phi_m p$.  Following
the same calculational steps as before, we now find
\be
\fft{\eta}{S} =
\fft{1}{4\pi\, TS} \Big(\fft32 M_{\rm grav} - \Phi_e Q_e - \Phi_m Q_m\Big)\,.
\ee
The generalized Smarr formula (\ref{n4smarr})
then implies that $\eta/S$ is again
given by (\ref{ratio}).

\subsection{Dyonic Kaluza-Klein AdS black hole}

   An explicit four-dimensional dyonic black hole solution was
found in \cite{gaugedyon}.  The solution has three non-trivial
parameters, which may
be viewed as the mass, the electric and the magnetic charges.  Although
it does not contain the full complement of four independent parameters of
the general solutions (which can only be obtained numerically), it
has sufficiently many parameters that the role of the ``charge''
for the scalar field can be exhibited in a non-trivial way.

The four-dimensional Lagrangian for this example is a consistent
truncation of ${\cal N}=8$ gauged supergravity (in fact, a consistent
truncation of the gauged STU model, where only one of the four $U(1)$
gauge fields is retained, and the axion in then consistently truncated also).
The Lagrangian
 is given by (\ref{emdlag}) with
\be
Z=e^{-\sqrt3\phi}\,,\qquad V=-6g^2 \cosh (\ft1{\sqrt3} \phi)\,.
\ee
Both the spherically-symmetric and planar asymptotic-AdS dyonic black
holes were constructed in \cite{gaugedyon}.  For our present purposes, we
shall consider just the planar black hole, which is given by \cite{gaugedyon}
\begin{eqnarray}
ds^2 &=& -(H_1 H_2)^{-\fft12} f dt^2 + (H_1 H_2)^{\fft12}
  \Big(\fft{d\rho^2}{f}
+ \rho^2 (dx^2 + dy^2)\Big)\,,\cr
\phi&=&\fft{\sqrt{3}}{2} \log\fft{H_2}{H_1}\,,\qquad
f=- \fft{\mu}{\rho} + g^2 \rho^2 H_1 H_2\,,\cr
A&=&\sqrt{\ft12{\mu}} \Big(\fft{(\rho + 2\beta_1)}{\sqrt{\beta_1}\,
H_1\,\rho}\, dt
+ 2\sqrt{\beta_2}\, xdy\Big)\,,\cr
H_1&=&1 + \fft{4\beta_1}{\rho} + \fft{4\beta_1\beta_2}{\rho^2}\,,\qquad
H_2=1 +\fft{4\beta_2}{\rho} + \fft{4\beta_1\beta_2}{\rho^2}\,.
\label{adsdyon2}
\end{eqnarray}
(We have rescaled $\mu$ by a factor of 2 relative to the solution presented
in \cite{gaugedyon}, to fit with our conventions in the rest of this
paper.)  Note that the radial coordinate being used here, which we call $\rho$
to distinguish it from $r$ that we are using in the rest of this paper,
is related to $r$ by $r= \rho\, (H_1\, H_2)^{1/4}$.

The parameter $\mu$ can be expressed in terms of the horizon
radius $\rho=\rho_0$, namely
\begin{equation}
\mu= g^2 \rho_0^3 H_1(\rho_0) H_2(\rho_0)\,,
\end{equation}
Here we are, for convenience, assuming that the
$\R^2$ coordinates $(x,y)$ have been identified to give a 2-torus of
volume $4\pi$.  One can take any other choice for the volume, with the
understanding that the extensive quantities should be scaled by the
relative volume factor. We find that the remaining thermodynamic
quantities are given by
\begin{eqnarray}
&&M=\fft{\mu}{8\pi}\,,\qquad Q_e=\fft1{8\pi}\sqrt{\mu\beta_1}\,,\qquad
Q_m=\fft1{8\pi}\sqrt{\mu\beta_2}\,,\cr
&&\Phi_e=\fft{2\sqrt{\mu\beta_1} (r_0 + 2\beta_2)}{r_0^2 H_1(r_0)}\,,\qquad
\Phi_m=\fft{2\sqrt{\mu\beta_2} (r_0 + 2\beta_1)}{r_0^2 H_2(r_0)}\,,\cr
\phi_1&=&2\sqrt3 (\beta_2-\beta_1)\,,\qquad
\phi_2=2\sqrt3 (\beta_1^2 - \beta_2^2)\,.
\end{eqnarray}
Note that $\phi_1$ and $\phi_2$ here are, as in the rest of this paper,
the leading coefficients of the large-distance expansion for $\phi$, using
the $r$ coordinate. Thus $\phi= \phi_1/r + \phi_2/r^2+\cdots$ here.
It is now straightforward to verify that
\be
dM=TdS + \Phi_e dQ_e + \Phi_m dQ_m +
  \fft{g^2}{48\pi} (\phi_1 d\phi_2 - 2\phi_2 d\phi_1)\,.\label{FLdyon}
\ee
It is worth noting that
\be
\phi_1\, d\phi_2 - 2 \phi_2\, d\phi_1 = 24(\beta_1-\beta_2)\,
  (\beta_2\, d\beta_1 - \beta_1\, d\beta_2)\,,
\ee
and so unlike in some simple solutions that have been discussed elsewhere
in the literature, here the scalar terms
$\phi_1\, d\phi_2 - 2 \phi_2\, d\phi_1$
play an essential role in the relation (\ref{FLdyon}) between the
infinitesimal variations of the black-hole parameters.  (The right-hand side
of (\ref{FLdyon}) would not be an exact form if these terms were omitted.)
It can easily be seen that the quantities given above obey the
 generalized Smarr formula (\ref{n4smarr}).


\section{Non-Minimally Coupled Scalar}

In this section, we consider Einstein-Maxwell-Dilaton theories in which
the dilaton couples non-minimally to gravity, with a Lagrangian given by
\be
e^{-1}{\cal L} = \kappa(\phi) R- \ft12 (\partial\phi)^2 -
     \ft14 Z(\phi) F^2 - V(\phi)\,,
\label{emdlag2}
\ee
for which the equations of motion are
\bea
\kappa(\phi)\, G_{\mu\nu} &=& \nabla_\mu\nabla_\nu\, \kappa(\phi) -
   \square\kappa(\phi)\, g_{\mu\nu} +\ft12\Big(\del_\mu\phi\, \del_\nu\phi
-\ft12(\del\phi)^2\, g_{\mu\nu}\Big)
  +\ft12 V(\phi)\, g_{\mu\nu} \nn\\
&& + \ft12 Z(\phi)\Big(F_{\mu\rho}\, F_\nu{}^\rho
-\ft14 F^2\, g_{\mu\nu}\Big)\,,\nn\\
\square\phi &=& \fft{\del V(\phi)}{\del\phi}
    -\fft{\del\kappa(\phi)}{\del\phi}\, R +
   \ft14 \fft{\del Z(\phi)}{\del\phi}\, F^2\,,\qquad \nabla_\mu\Big(Z(\phi)\, F^{\mu\nu}\Big) =0\,,
\eea
where $G_{\mu\nu}=R_{\mu\nu}-\ft12 R\, g_{\mu\nu}$ is the Einstein
tensor.
We shall assume that both the functions $\kappa(\phi)$ and $Z(\phi)$
become unity when $\phi=0$, which is a stationary point of the potential
$V(\phi)$.

    We consider black holes of the type (\ref{chargedbb}), with an
horizon located at $r=r_0$.  We may easily verify that the linearised
transverse and traceless perturbative mode is again given by
(\ref{chargedpsi}), but with the function $U$ now given by
\be
U=\fft{q}{\sqrt{h_1 f_1}} \int \fft{a}{r^{n-2} \kappa(\phi)\sqrt{hf}}\,,
\ee
where $a$ is chosen to be in the gauge for which it vanishes on the horizon.
Since now the relevant surface term in the quadratic
action of the linearised mode is
\be
-\ft12\, r^{n-2}\, \kappa(\phi) \sqrt{fh}\,
\Psi\Psi'\Big|_{r=\infty} \,,\label{surfaction2}
\ee
rather than the previous expression (\ref{surfaction1}), it follows that
the viscosity/entropy ratio will be the same as that given in
(\ref{chargedetas}), where now the entropy is given by
$\kappa (\phi)|_{r=r_0}$ multiplying the area of the horizon.  Thus we
see that again $\eta/S= 1/(4\pi)$, as a consequence of the
generalised Smarr formula.

Many examples of exact black-hole solutions in such theories have
been found in the literature
\cite{Martinez:1996gn,Nadalini:2007qi,Anabalon:2012ta,Zou:2014gla,Fan:2015tua}
and it can easily be verified that all of them indeed satisfy the
generalised Smarr formula.  Of course, these exact solutions all have
fewer than the maximal number of independent parameters,
and the first law of thermodynamics therefore does not involve the
scalar charges.  The working of the generalised Smarr formula, and hence
the saturation of the viscosity bound,  is therefore less striking in these
explicit special cases.

\section{Noether Charge, Generalised Smarr formula and Viscosity Bound}

In the previous sections, we established the link between the saturation
of the viscosity bound and the generalised Smarr relation in the
general Einstein-Maxwell-Dilaton theory, both with a minimal and with a
non-minimal
scalar coupling to gravity.  Owing to the existence of the extra scalar
``charges'' in asymptotically-AdS spacetimes, one might have
expected the generalised Smarr
formula to include a contribution from these, but as we saw in detail
this does not occur.  We derived this result by making use of scaling arguments
and the first law of black  hole (thermo)dynamics, derived from
the Wald formalism.  Although instructive as a demonstration of
the interplay between scaling and the first law, the derivation was
somewhat indirect.  In this
section, we obtain further insights by presenting
a different derivation of the generalised
Smarr relation, based on the construction of a Noether charge
associated with the
relevant scaling symmetry.

\subsection{Einstein-Maxwell-Dilaton theory}

We start with the most general theory (\ref{emdlag2}) we have considered
in this paper, which encompasses all the previous cases.  We rewrite
the black hole ansatz as
\be
ds^2 = - u\, dt^2 + d\rho^2 + v\, dx^i dx^i\,,\qquad
A=a\, dt\,,
\ee
where $u$, $v$, $a$ and the scalar $\phi$ are all functions only of
the coordinate $\rho$.  Since this is the most general ansatz that
respects the isometries, we can safely substitute the ansatz into the
Lagrangian, finding
\be
{\cal L} =u\,v^{n-2} \Big[\kappa\big(-2\fft{\ddot u}{u} -2(n-2)\fft{\ddot v}{v} -
  (n-2)(n-3)
\fft{\dot v^2}{v^2} - 2(n-2) \fft{\dot u \dot v}{u v}\big) -
\ft12 \dot \phi^2 -\ft12
Z \fft{\dot a^2}{u^2} - V\Big]\,,
\ee
where a dot denotes a derivative with respect to $\rho$.  The Lagrangian
is invariant under the global scaling
\be
u\rightarrow \lambda^{-(n-2)} \,u\,,\qquad v\rightarrow \lambda \,v\,,
\qquad a\rightarrow \lambda^{-(n-2)} \, a\,.
\ee

  If we now allow $\lambda$ to be $\rho$ dependent, then by integrating by
parts and collecting the coefficient of $\dot \lambda$, we can derive the
conserved Noether charge
\be
2\kappa (v \dot u -u\dot v) v^{n-3} -Z a u^{-1} v^{n-2}\,
\dot a =c=\, {\rm const.}
\ee
Re-writing this equation using the $r$ coordinate defined in
(\ref{chargedbb}), and substituting also (\ref{aprime}), we find
\be
\kappa\sqrt{hf}\,  \Big(\fft{h'}{h} -
\fft{2}{r}\Big) \, r^{n-2} + q\,a= c\,.
\label{firstorder}
\ee
Note that for $q=0$ and $\kappa(\phi)=1$, this is just
the first integral of the second-order differential equation (\ref{ESanseom2}).

   Assuming that the solution is asymptotic to AdS, for which it is
necessary (but not sufficient) that $f$ and $h$ have the asymptotic forms
$h =g^2 r^2+ \cdots$ and $f=g^2 r^2 + \cdots$ as
$r\rightarrow \infty$, and assuming also
that $\kappa\rightarrow 1$ asymptotically, we can then
conclude from (\ref{firstorder}) that
\be
h=g^2r^2 - \fft{\mu}{r^{n-3}} + \cdots\,,
\ee
with no slower-falling intervening terms between the $g^2 r^2$
and the $-\mu/r^{n-3}$ terms.\footnote{We emphasise that this absence of
intervening terms in the asymptotic form of $h$ is deduced using the
first-order equation (\ref{firstorder}), which is valid for
planar black holes but not spherically-symmetric black holes.  Indeed, in
our discussion in section 3.4 we discussed cases where the
spherically-symmetric black-hole solutions presented in \cite{lupowe}
had such intervening terms in the asymptotic expansion of $h$, but these
all scaled away when the planar limit was taken.}
Correspondingly, the constant $c$ in (\ref{firstorder}) is simply
given by $c=(n-1)\mu$.
Applying the identity (\ref{firstorder}) instead on the horizon, we therefore
find
\be
(n-1)\mu = 16\pi (T S + \Phi_e Q_e)\,.\label{gensmarrnc}
\ee
It then follows from (\ref{Mgravdef}) that generalized Small formula
(\ref{smarrPhi})
holds (or, with a straightforward extension of the above argument in
$n=4$ dimensions, (\ref{n4smarr})), and hence the viscosity
bound $\eta/S\ge 1/(4\pi)$ is saturated.  Note that in this calculation,
we have chosen the gauge where $a$ vanishes at infinity. Alternatively,
we could choose the gauge where $a$ vanishes instead on the horizon.
In this case, the constant $c$ becomes $(n-1)\mu - 16\pi \Phi_e Q_e$,
when evaluated at asymptotic infinity, and $16\pi TS$ when evaluated
on the horizon.  The generalized Smarr relation (\ref{gensmarrnc})
continues to hold.

\subsection{Gauss-Bonnet gravity}

We now examine the relation between the
generalised Smarr formula and the viscosity/entropy ratio
for an example of a theory with
higher-derivative terms, namely Einstein-Maxwell theory with a
cosmological constant and an
added Gauss-Bonnet term.  We take the Lagrangian in general
dimensions to be given by
\be
e^{-1} {\cal L}_n = R -
\alpha (R^2 - 4 R^{\mu\nu}R_{\mu\nu} + R^{\mu\nu\rho\sigma}
    R_{\mu\nu\rho\sigma}) - \ft14 F^2 + (n-1)(n-2) g^2\,.
\ee
The exact solution for the planar black hole in Gauss-Bonnet gravity was
constructed in \cite{cai}. In fact, however, in the following
discussion it will not be necessary to know the explicit form of the
solution.
Following the same procedure as in the previous subsection, we obtain the
Noether charge
\be
\sqrt{hf} \,
\Big(\fft{h'}{h} - \fft{2}{r}\Big) \Big(r^2-2(n-3)(n-4)\alpha f\Big)
\, r^{n-4}+ q\, a= c\,.\label{gbfirstorder}
\ee

  In dimensions $n\ge5$ the Gauss-Bonnet term modifies the effective
cosmological constant in asymptotically-AdS solutions, such that
$h$ and $f$ will have the leading-order asymptotic forms $h= \tilde g^2 r^2
+\cdots$, $f=\tilde r^2 + \cdots$, where $g$ and $\tilde g$ are related by
\be
 g^2 = \tilde g^2\, [1- (n-3)(n-4)\, \alpha\, \tilde g^2]\,.
\ee
By evaluating (\ref{gbfirstorder}) asymptotically and on the horizon,
in the same manner is in the previous subsection, it is then straightforward
to see that $c=(n-1)\mu\, [1-2\alpha (n-3)(n-4)\, \tilde g^2]$ and hence
we obtain a modification to the previous generalised Smarr formula, with
\be
(n-1) \, [1-2\alpha (n-3)(n-4)\, \tilde g^2]\, \mu =
     16\pi (T S + \Phi_e Q_e)\,.
\ee
Since the mass is given by
\be
M = \fft1{16\pi}\, (n-2)\, [1-2\alpha\, (n-3)(n-4)\,\tilde g^2]\, \mu\,,
\ee
the usual generalised Smarr relation (\ref{smarrPhi}) continues to hold.
However, as shown in \cite{shenker} for a neutral planar black hole in
$n=5$ dimensions,
the viscosity/entropy ratio, after
re-expressing in our notation, is given by
\be
\fft{\eta}{S}= \fft1{4\pi}\, (1-4\alpha\, \tilde g^2)^2\,,
\ee
and so the correlation between the Smarr relation and the viscosity bound
no longer holds in this higher-derivative example.

\section{Conclusions}

  Motivated by an interest in trying to understand the rather
widespread universality of the ratio $\eta/S=1/(4\pi)$ in boundary field
theories dual to bulk two-derivative theories involving gravity,
we have investigated the relation to a generalisation of the Smarr formula
of classical general relativity.  The generalised Smarr formula in
question arises as
a consequence of a scaling symmetry that is a specific feature of planar
black holes, and which is absent in the case of black holes with
spherical symmetry.

   One way in which the generalised Smarr relation can be derived is
via thermodynamic considerations.  Starting from the first law of
thermodynamics for a class of black-hole solutions, and given a scaling
symmetry of the system of solutions, one can essentially integrate up
the first law to obtain an algebraic expression for the black hole mass
as a sum of products of the thermodynamic quantities characterising the
solutions times their thermodynamic conjugate variables.  Our principal
focus in this paper has been to study the resulting generalised
Smarr formulae in cases that have not been extensively studied
previously, in which a scalar field plays an essential role and in fact
leads to the enlargement of the parameter-space of black hole solutions.
In the simplest example, of black holes in Einstein gravity minimally
coupled to a scalar field with an appropriate potential, the symmetrical
black holes we consider depend not merely on a single mass parameter
(as in Schwarzschild-AdS) but on two parameters, which can be thought of
as the mass and a scalar ``charge'' in addition.

  We have shown in this paper that although the scalar charge and its
thermodynamic conjugate variable enter non-trivially in the first law (in fact,
precisely {\it because} they enter in the first law), they
do not contribute in the generalised Smarr relation, which continues to
take the same form (\ref{smarrM}) as in the simple case when the
scalar field vanishes.  The formula (\ref{smarrM}) provides an
exact expression for the gravitational mass of the black hole, which is
a quantity associated with a parameter in the asymptotic solution at
infinity, in terms of the quantities $T$ and $S$, which are defined at
the black-hole horizon.  Thus even though the two-parameter black-hole
solutions cannot be found explicitly, owing to the complexity of the
equations of motion for the system, one still has an exact result for the
ratio of mass to $TS$.  As we have seen, this ratio is proportional
to the viscosity to entropy ratio $\eta/S$, and so we are able to give an
exact evaluation of this ratio, leading to the familiar result $1/(4\pi)$,
even in this rather complicated class of examples.

   One of the goals of this paper has been to derive the generalised
Smarr relation for planar black holes in theories involving a scalar
field by making use of the first law of thermodynamics, precisely
because of the subtleties that arise in the first law in this case.
As we saw, although the coefficients $\phi_1$ and $\phi_2$ in the
asymptotic expansion of the scalar field enter non-trivially in the
first law, the term proportional to $\phi_1\, \phi_2$ that one
might have expected, \`a priori, to enter as a contribution in the
generalised Smarr relation is actually absent. Thus by making use of the
first law (\ref{FL0}), we derived the generalised Smarr formula
(\ref{smarrM}) that relates the horizon quantities $T$ and $S$ to
the asymptotic quantity $M_{\rm grav}$ (which is proportional to the
coefficient $\mu$ in the asymptotic expansion of $h(r)$).  If one
had taken a more restrictive viewpoint in which $\phi_1$ and $\phi_2$
were held fixed, or a functional relation between them imposed, then
the first law would simply have read $dM=TdS$, the scaling symmetry
would have been broken, and one would not have been able to derive
an expression for $\mu$ in terms of $T$ and $S$ by this method.

   The generalised Smarr relation can also be derived in a
different way, by using the scaling symmetry of the planar black-hole
solutions to derive a conserved Noether charge, of the form (\ref{firstorder}).
In fact this expression not merely gives an equality of the
the left-hand side at infinity (proportional to the mass) to the left-hand
side on the horizon, but an equality valid at all radii $r$.  This
alternative derivation of the generalised Smarr formula also provides
a further vindication of the thermodynamic interpretation
\cite{gaugedyon,Liu:2013gja,lupowe}  that we have
adopted.

   We also extended our discussion to a wide class of Einstein-Maxwell-Dilaton
theories, with Lagrangian given by (\ref{emdlag2}).  These theories
in general have non-minimal coupling of the scalar field to gravity,
with minimal coupling arising if $\kappa(\phi)=1$.   We showed that a
generalised Smarr relation of the form (\ref{smarrPhi}) (or (\ref{n4smarr}) in
$n=4$ dimensions) holds for
the general planar black-hole solutions in all these theories. Furthermore,
we showed also that $\eta/S$ again equals $1/(4\pi)$ for all the
black-hole solutions.  Note that the proof was general, in the
sense that it did not require knowledge of the explicit form of the
black-hole solutions.

   We have seen that the equality $\eta/S=1/(4\pi)$ and the
generalised Smarr relation go hand-in-hand for the rather general class of
two-derivative theories that we have investigated.  One might, in fact, say
that the generalised Smarr formula is the bulk gravity holographic dual
of the saturation of the $\eta/S$ bound in the boundary field theory.  This
mapping breaks down in cases such as the higher-derivative theory involving
the Gauss-Bonnet term.  It would be interesting to try to obtain a deeper
understanding of the circumstances under which the breakdown should occur.

\section*{Acknowledgements}

H-S.L.~is supported in part by NSFC grant 11305140, SFZJED grant Y201329687 and CSC scholarship No. 201408330017. C.N.P.~is supported in part by DOE grant DE-FG02-13ER42020. The work of H.L.~is supported in part by NSFC grants NO. 11175269, NO. 11475024 and NO. 11235003.

\appendix

\section{Dimensional Scaling and the Standard Smarr Relation}

  In this appendix, we present a discussion of the standard Smarr
formula for black holes in theories of gravity, as a contrast to the
generalised Smarr formulae that have formed the focus of the rest of
the paper.

  One can obtain a relation of the general form of a Smarr relation
whenever a solution has a scaling symmetry.  The classic Smarr relation
holds in the case of asymptotically-flat black hole solutions, with
the scaling symmetry being the one where all parameters in the theory
simply scale according to their ``engineering'' dimensions.  For example,
in the Reissner-Nordstr\"om black hole in $n$ dimensions, one
has
\be
ds^2 = -h dt^2 + \fft{dr^2}{h} + r^2\, d\Omega_{n-2}^2\,,\qquad
h= 1 - \fft{\mu}{r^{n-3}} + \fft{q^2}{r^{2n-6}}\,,\label{RNn}
\ee
and the first law of thermodynamics reads
\be
dM = TdS + \Phi\, dQ\,,\label{RNFL}
\ee
where $\Phi$ is the potential difference between the horizon and infinity,
and $Q$ is the conserved charge.  From the definition of $h$ in (\ref{RNn})
we see that $[\mu]=L^{n-3}$ and $[q]=L^{n-3}$.  The canonical conserved
mass $M$ and charge $Q$ have the same dimensions as $\mu$ and $q$ respectively,
and hence there is a scaling symmetry with
\be
M=\lambda^{n-3}\, \widetilde M\,,\qquad Q=\lambda^{n-3}\, \widetilde Q\,,
\qquad \Phi=\widetilde\Phi\,,\qquad T= \lambda^{-1}\, \widetilde T\,,
\qquad S= \lambda^{n-2}\, \widetilde S\,.\label{standardscale}
\ee
Thus $M(\lambda^{n-2}\, \widetilde S,\lambda^{n-3}\, \widetilde Q)=
   \lambda^{n-3}\, \widetilde M(\widetilde S,\widetilde Q)$, and
so acting with $\lambda\, \del/\del\lambda$ and using (\ref{RNFL}),
one obtains the standard Smarr relation
\be
M= \fft{(n-2)}{(n-3)}\, TS + \Phi\, Q\,.\label{standardsmarr}
\ee

   If we now consider the Reissner-Nordstr\"om AdS black hole, where
a cosmological constant $\Lambda=- (n-1)\, g^2$ has been added to the
theory, and the metric function $h$ in (\ref{RNn}) becomes
\be
h= g^2 r^2 + 1 - \fft{\mu}{r^{n-3}} + \fft{q^2}{r^{2n-6}}\,,
\ee
then the scaling symmetry is broken if $\Lambda$, which has the dimension
$[\Lambda]=L^{-2}$, is treated conventionally as a fixed parameter in the
theory.  As a consequence, one no longer has a Smarr relation of the
form (\ref{standardsmarr}).

   A Smarr relation for Reissner-Nordstr\"om-AdS black hole {\it can} be
obtained if the viewpoint is changed slightly and $\Lambda$, despite
being a parameter in the Lagrangian, is treated as if it were a thermodynamical
variable.  The first law (\ref{RNFL}) then generalises to
\be
dM = TdS + \Phi\, dQ + \Upsilon\, d\Lambda\,,\label{RNFLLam}
\ee
where, since the cosmological constant acts like a pressure, its
thermodynamic conjugate $\Upsilon$ is like a volume.  The standard
dimensional scalings are then augmented by
\be
\Lambda = \lambda^{-2}\, \widetilde\Lambda\,,\qquad
\Upsilon = \lambda^{n-1}\, \widetilde\Upsilon\,,
\ee
and the scaling relation becomes
$M(\lambda^{n-2}\, \widetilde S,\lambda^{n-3}\, \widetilde Q,
\lambda^{-2}\, \widetilde\Lambda)=
   \lambda^{n-3}\, \widetilde M(\widetilde S,\widetilde Q,\widetilde\Lambda)$.
Acting with $\lambda\, \del/\del\lambda$ and using (\ref{RNFLLam}), one
obtains the Smarr relation
\be
M= \fft{(n-2)}{(n-3)}\, TS + \Phi\, Q - \fft{2}{(n-2)}\, \Upsilon\, \Lambda
\,.\label{standardsmarrLam}
\ee
Although the idea of allowing $\Lambda$ to vary in the first law is not
part of the classic treatment of black hole thermodynamics, we may still
refer to (\ref{standardsmarrLam}) as a ``standard'' type of Smarr relation
in the sense that it is based on the scaling symmetry that one can always
realise, in which all parameters (including parameters in the Lagrangian if
necessary) are scaled simply according to their ``engineering'' dimensions.
We illustrated the discussion above with the simple examples of the
Reissner-Nordstr\"om black holes, with or without cosmological
constant, but of course the whole discussion can be extended to include
all black holes, with rotation as well, and in more complicated theories
such as supergravities.

    Another example of a standard Smarr relation is for the
gauge dyonic black hole that we discussed in section 5.
As was shown in \cite{gaugedyon}), it admits a ``standard'' type of
Smarr relation (with the gauge parameter, related to $\Lambda$ by $\Lambda=
-3g^2$, treated as a thermodynamic variable too).  The first law reads
\be
dM=TdS + \Phi_e dQ_e + \Phi_m dQ_m + \Upsilon d\Lambda +
\fft{g^2}{48\pi} (\phi_1 d\phi_2 - 2\phi_2 d\phi_1)\,,
\ee
where
\be
\Upsilon = -\fft1{24\pi} \big(4 \beta_1 \beta_2 (\beta_1 + \beta_2) + 12 \beta_1 \beta_2 r_0 + 3 (\beta_1 + \beta_2) r_0^2 + r_0^3\big)\,.
\ee
It is then clear that the first law is invariant if  all the thermodynamical
quantities are scaled according to their physical dimensions.  This leads
to the ``standard'' Smarr formula
\be
M=2 T S + \Phi_e Q_e + \Phi_m Q_m - 2\Upsilon \Lambda\,.
\ee
(This can be done both for the spherically-symmetric or the planar black holes,
but here we are presenting $\Upsilon$ just for the planar limit that is
relevant for our discussion in this paper.)

   All of the above discussion in this appendix was concerned with the
``standard'' Smarr relations that one can derive by considering a scaling
of quantities according to their ``engineering'' dimensions.  This is to
be contrasted with the scaling symmetry that has been the focus of our
attention in this paper, which applies only to planar black holes
and not to spherically-symmetric black holes.  In particular, the
scaling symmetry we have been using is one where $r$ scales
as $r\rightarrow \lambda\, r$ but the mass scales as $M\rightarrow
\lambda^{n-1}\, M$, and {\it not} as $M\rightarrow \lambda^{n-3}\, M$
as it does in dimensional scaling (see (\ref{standardscale})).  This is
because the metric function $h$ is of the form
\be
h= g^2 r^2 -\fft{\mu}{r^{n-3}} +\cdots
\ee
in the planar black holes, rather than $h= g^2 r^2 + 1 -\mu/r^{n-3} +\cdots$
in spherically-symmetric black holes. The scaling symmetry we wish to
consider in this paper is one where $g$ is held fixed, and thus in order for
the $g^2 r^2$ and the $-\mu/r^{n-3}$ terms in $h$ to scale the same way,
we must have $\mu\rightarrow \lambda^{n-1}\, \mu$.  This scaling
symmetry would be broken if the ``1'' term were present in $h$ and
$f$, as it is in the spherically-symmetric black holes.  We have consistently
referred to Smarr relations derived from the $\mu\rightarrow \lambda^{n-1}\,
\mu$ scaling as ``generalised,'' to contrast them with the ``standard''
relations based on the $\mu\rightarrow \lambda^{n-3}\, \mu$ dimensional
scaling.

\end{document}